\documentclass[aps,prl,twocolumn,showpacs,preprintnumbers,nofootinbib,float,longbibliography,superscriptaddress]{revtex4-1}

\usepackage{epsfig}
\usepackage[dvipsnames]{xcolor}
\usepackage{float,amsmath,amssymb}
\usepackage{graphicx}
\usepackage{url}
\usepackage{bbold}
\usepackage[utf8]{inputenc}
\usepackage{physics}

\newcommand{\xpom}{x_\mathbb{P}}

\newcommand{\rt}{{\mathbf{r}_\perp}}

\newcommand{\xt}{{\mathbf{x}_\perp}}

\newcommand{\bt}{{\mathbf{b}_\perp}}
\newcommand{\bti}{{\mathbf{b}_{\perp,i}}}
\newcommand{\Deltat}{{\boldsymbol{\Delta}_\perp}}

\newcommand{\nc}{{N_\mathrm{c}}}
\newcommand{\jpsi}{$\mathrm{J}/\psi$ }
\newcommand{\jpsim}{\mathrm{J}/\psi}

\usepackage[breaklinks,colorlinks,citecolor=citcolor,urlcolor=blue,linkcolor=lcolor]{hyperref}
\definecolor{lcolor}{rgb}{0.5,0,0}
\definecolor{citcolor}{rgb}{0,0.3,0.0}

\begin{document}

\title{Multi-scale Imaging of Nuclear Deformation at the Electron Ion Collider}

\author{Heikki M\"antysaari}
\affiliation{Department of Physics, University of Jyv\"askyl\"a, P.O. Box 35, 40014 University of Jyv\"askyl\"a, Finland}
\affiliation{Helsinki Institute of Physics, P.O. Box 64, 00014 University of Helsinki, Finland}

\author{Bj\"orn Schenke}
\affiliation{Physics Department, Brookhaven National Laboratory, Upton, NY 11973, USA}

\author{Chun Shen}
\affiliation{Department of Physics and Astronomy, Wayne State University, Detroit, Michigan 48201, USA}
\affiliation{RIKEN BNL Research Center, Brookhaven National Laboratory, Upton, NY 11973, USA}

\author{Wenbin Zhao}
\affiliation{Department of Physics and Astronomy, Wayne State University, Detroit, Michigan 48201, USA}

\begin{abstract}
We show within the Color Glass Condensate framework that exclusive vector meson production at high energy is sensitive to the geometric deformation of the target nucleus at multiple length scales. Studying $e+$U collisions and varying the deformation of the uranium target, we demonstrate that larger deformations result in enhanced incoherent vector meson production cross sections. Further, different multipole deformation parameters affect different regions of transverse momentum transfer. Employing JIMWLK evolution to study the Bjorken-$x$ dependence of our results, we find that the ratio of incoherent to coherent cross sections decreases with decreasing $x$, largely independently of the quadrupole deformation of the target. Comparing results for the same process using ${\rm {^{20}Ne}}$ targets with ${\rm {^{16}O}}$ targets,
we find that differences in deformation are clearly visible in the incoherent cross section. 
These findings show that certain observables at the Electron-Ion Collider are very sensitive to nuclear structure. Consequently, deformations need to be taken into account when interpreting experimental results. More importantly, this also means that  $\vert t\vert$-differential diffractive vector meson production could become a powerful tool, enabling the most direct measurements of nuclear structure at different length scales, ranging from nuclear deformation at low $\vert t\vert$ to nucleon- and subnucleon-size scales at higher $\vert t\vert$.
\end{abstract}

\maketitle

\noindent{\it 1. Introduction.}
Understanding the geometric structures of the proton and nuclei, including their event-by-event fluctuations, is of fundamental interest. Deep inelastic scattering (DIS) of leptons on hadrons is one of the most important tools to probe the partonic structure of protons and nuclei.  
In the 2030's, the Electron-Ion Collider in the US will provide access to nuclear-DIS for the first time in collider kinematics~\cite{AbdulKhalek:2021gbh,Aschenauer:2017jsk}. In addition to the EIC, there are also other longer-term proposals for future nuclear-DIS facilities at CERN~\cite{LHeC:2020van} and in China~\cite{Anderle:2021wcy}.

Exclusive vector meson (e.g.~$\mathrm{J}/\psi$) production off nucleons and nuclei is a particularly  clean and powerful process to probe the nuclear high-energy structure at small longitudinal momentum fraction $x$ for several reasons.
First,  in order to produce only the vector meson and nothing else, net color charge cannot be transferred to the target, and the final state is unambiguously identified by a large rapidity gap. This also requires that at least two gluons are exchanged, which renders the cross section approximately proportional to the squared gluon distribution~\cite{Ryskin:1992ui} at leading order (see Ref.~\cite{Eskola:2022vpi} for a recent analysis at next-to-leading order accuracy). Additionally, only in such exclusive scattering  it is possible to determine the total momentum transfer to the target hadron or nucleus, which is the Fourier conjugate to the impact parameter and as such provides access to the target geometry.

In the hot QCD community, understanding the spatial shape of colliding objects in heavy-ion collisions is a necessary input when the deconfined  Quark-Gluon Plasma (QGP) state is probed. The initial nucleon and nuclear geometry determine the spatial distribution of nuclear matter in the collision, which in turn determines the initial pressure anisotropies that are transformed into observable momentum space correlations
when the space-time evolution of the QGP is simulated. 
The hydrodynamic modeling of heavy ion collisions with an accurate initial state description in collisions of several different ion species has revealed how the detailed structural properties of the colliding nuclei are visible in final state particle correlations~\cite{Filip:2009zz,Masui:2009qk,Hirano:2012kj,Shen:2014vra,Schenke:2014tga,Schenke:2020mbo,Xu:2021uar,Giacalone:2021udy, Zhang:2021kxj,Nijs:2021kvn,Cheng:2023ciy, Ryssens:2023fkv}. 

The study of collisions involving heavy ions at the Relativistic Heavy Ion Collider (RHIC), the Large Hadron Collider (LHC), and the future Electron-Ion Collider (EIC) offers a synergistic approach to understanding the properties of high energy nuclei and hot and dense nuclear matter. The EIC will probe in detail the small-$x$ structure of the nuclei, obtaining fundamentally interesting information and constraining the initial state of heavy ion collisions \cite{Mantysaari:2016jaz,Mantysaari:2016ykx,Mantysaari:2020axf}. In turn, as mentioned above, heavy ion collisions can themselves constrain the high-energy structure of nuclei, including deformation and nucleon clustering.
The complementary information from these facilities is essential for gaining a complete picture of the properties of the QGP, and the influence of nuclear geometry in the initial state. 
 
We employ the Color Glass Condensate (CGC) framework \cite{Gelis:2010nm} supplemented with a model that describes the nuclear geometry including nucleon substructure in terms of gluonic hot spots~\cite{Mantysaari:2020axf}. Recently, in Ref.~\cite{Mantysaari:2022ffw}, we performed a statistically rigorous Bayesian analysis to extract the posterior likelihood distribution for the model parameters describing the event-by-event fluctuating proton geometry from HERA \jpsi production data~\cite{H1:2013okq}.
In this work, we extend our studies to $e+A$ collisions and for the first time explore the effects of nuclear deformation on the diffractive cross sections for large nuclei.
Specifically, we study the dependence of the $\vert t\vert$ differential diffractive \jpsi cross sections on the deformation parameters of uranium. To examine how energy (or Bjorken-$x$) evolution modifies the effects of deformation, we perform numerical simulations solving the JIMWLK equations (see e.g.~\cite{Mueller:2001uk}) to evolve the nuclear configurations to smaller Bjorken-$x$. 

We further study vector meson production in electron scattering off smaller nuclei. Of particular interest are ${\rm{^{20} Ne}}$ and ${\rm{^{16}O}}$, which have a similar mass number, but are expected to differ in shape. Calculations in different models~\cite{Ebran:2012ww,Zhou:2015nza,Marevic:2018crl,Frosini:2021sxj,Bally:2023d} obtain a characteristic bowling pin shape for ${\rm{^{20} Ne}}$, essentially forming a ${\rm{^{16}O}}$ like structure with an additional $\alpha$ cluster in its periphery.
We provide model predictions for future EIC measurements, in particular for the ratio of the $|t|$ differential diffractive \jpsi production cross sections in the two systems.
  
\bigskip
\noindent {\it 2. Vector meson production at high energy.}
The total diffractive cross section in DIS  gives insight into the total small-$x$ gluon densities of the target nuclei. More differential observables such as the exclusive production of a vector meson, $\gamma^* + p/A \to V + p/A $,  as a function of (squared) momentum transfer $-t$ can provide more detailed information on the target structure. 
The coherent cross section, corresponding to the process where the target remains in the same quantum state, can be obtained by averaging over the target color charge configurations $\Omega$ at the amplitude level~\cite{Good:1960ba}:
\begin{equation}
\label{eq:coherent}
     \frac{\dd \sigma^{\gamma^* + A \to V + A}}{\dd |t|}  = \frac{1}{16\pi} \left|\left\langle \mathcal{A} \right\rangle_\Omega\right|^2.
\end{equation}
The incoherent vector meson production cross section, for which the final state of the target is different from its initial state, is obtained by subtracting the coherent contribution from the total diffractive vector meson production cross section~\cite{Miettinen:1978jb,Caldwell:2010zza,Mantysaari:2020axf}. 
The incoherent cross section thus has the form of a variance
\begin{multline}
\label{eq:incoherent}
     \frac{\dd \sigma^{\gamma^* + A \to V + A^*}}{\dd |t|}  = \frac{1}{16\pi} \left[
     \left\langle \left|\mathcal{A}\right|^2\right \rangle_\Omega \right. 
     - \left. \left|\left\langle \mathcal{A}\right\rangle_\Omega\right|^2 \right]\,.
\end{multline}
Here $\mathcal{A}$ is the scattering amplitude for diffractive vector meson production, which at high energy describes the splitting of the virtual photon into a quark anti-quark pair, the pair's subsequent interaction with the target, followed by the formation of the vector meson. It can be written as~\cite{Kowalski:2006hc,Hatta:2017cte} (see also Refs.~\cite{Mantysaari:2022kdm,Mantysaari:2021ryb} for recent developments towards NLO accuracy)
\begin{multline}
\label{eq:jpsi_amp}
    \mathcal{A} = 2i\int \dd[2]{\rt} \dd[2]{\bt}  \frac{\dd{z}}{4\pi} e^{-i \left[\bt - \left(\frac{1}{2}-z\right)\rt\right]\cdot \Deltat} \\
    \times [\Psi_V^* \Psi_\gamma](Q^2,\rt,z) N_\Omega(\rt,\bt,\xpom).
\end{multline}
Here $\rt$ is the transverse size of the $q\bar q$ dipole,  $\bt$ is the impact parameter measured relative to the target center, and $Q^2$ is the photon virtuality. The fraction of the large photon plus-momentum carried by the quark is given by $z$, $\xpom$ is the fraction of the target longitudinal momentum transferred to the meson in the frame where the target has a large momentum, and $\Deltat$ is the transverse momentum transfer, with $-t \approx {\boldsymbol \Delta}_\perp^2$. The $\gamma^* \to q\bar q$ splitting is described by the virtual photon light front wave function $\Psi_\gamma$~\cite{Kovchegov:2012mbw}. The vector meson wave function $\Psi_V$ is non-perturbative and needs to be modeled, introducing some uncertainty. Here, we use the Boosted Gaussian parametrization from~\cite{Kowalski:2006hc}, where the model parameters are constrained by the decay width data. 

Dependence on the small-$x$ structure of the target is included in the dipole amplitude $N_\Omega(\rt,\bt,\xpom)$, which, for a given target color charge configuration $\Omega$, is  
$N_{\Omega}(\rt,\bt,\xpom) =  1 - \frac{1}{\nc} \tr \left[ V\left(\bt + \frac{\rt}{2}\right) V^\dagger\left(\bt - \frac{\rt}{2}\right) \right].$
The  $V(\xt)$ represents a Wilson line, depending on $\Omega$ and $\xpom$, and describing the color rotation of a quark state when propagating through the target field at transverse coordinate $\xt$. 
The Wilson lines are obtained in the same way as in the IP-Glasma initial state description~\cite{Schenke:2012wb} used e.g.~in Refs.~\cite{Mantysaari:2022sux,Mantysaari:2020lhf,Mantysaari:2019jhh,Mantysaari:2019csc,Mantysaari:2018zdd,Mantysaari:2016jaz,Mantysaari:2016ykx}. 
They are computed by first relating the average square color charge density to the local saturation scale extracted from the IPSat dipole-proton amplitude~\cite{Schenke:2012hg}. Then, by solving the Yang-Mills equations for the gluon fields, one obtains
\begin{equation}
  V(\xt) = \mathrm{P}_{-}\left\{ \exp\left({-ig\int_{-\infty}^\infty \dd{z^{-}} \frac{\rho^a(x^-,\xt) t^a}{\boldsymbol{\nabla}^2 - m^2} }\right) \right\}\,,
  \label{eq:wline_regulated}
\end{equation}
where $\mathrm{P}_{-}$ represents path ordering in the $x^-$ direction and $\rho^a$ is the color charge density. Here, we introduced the infrared regulator $m$, which is needed to avoid the emergence of unphysical Coulomb tails. 

We note that in the incoherent cross section \eqref{eq:incoherent}, the square of the impact parameter dependent scattering amplitude is equivalent to the Fourier transform of the two-point function of the nuclear thickness function, which clarifies the sensitivity of this quantity to the target structure at different length scales, depending on $|t|$ (see also \cite{Caldwell:2010zza,Blaizot:2022bgd}).

In this work we use sub-nucleonic fluctuations of the nucleon, introducing an event-by-event fluctuating density by following Refs.~\cite{Mantysaari:2016jaz,Mantysaari:2016ykx} and  writing the density profile of nucleons $T_p(\bt)$ as
\begin{equation}
\label{eq:Tpfluct}
    T_p(\bt) = \frac{1}{N_q} \sum_{i=1 }^{N_q} p_i T_q(\bt-\bti),
\end{equation}
where the single hot spot density distribution $T_q(\bt) = \frac{1}{2\pi B_q} e^{-{\mathbf b}_\perp^2/(2B_q)}\,$ and the coefficient $p_i$ allows for different normalizations for individual hot spots. It follows the log-normal distribution with the width $\sigma$ controlling the magnitude of the density fluctuations. 
Our prescription corresponds to having $N_q$ hot spots with hot spot width $B_q$. The hot spot positions $\bti$ are sampled from a two-dimensional Gaussian distribution whose width is denoted by $B_{qc}$, and the center-of-mass is shifted to the origin at the end. In this work, we use the Maximum a Posteriori (MAP) parameter set from Bayesian analysis where the geometry parameters at $\xpom\approx 0.0017$ are constrained by the exclusive \jpsi production data from HERA~\cite{Mantysaari:2022ffw}.

To model the geometric shape of large nuclei, we first sample nucleon positions from a Woods-Saxon distribution 
\begin{equation}\label{eq:WS}
    \rho(r,\theta) = \frac{\rho_0}{1+\exp[(r-R'(\theta))/a]}\,,
\end{equation}
with $R'(\theta)=R[1+\beta_2 Y_2^0(\theta)+\beta_3 Y_3^0(\theta) +\beta_4 Y_4^0(\theta)]$, and $\rho_0$ is the nuclear density at the center of the nucleus. Here $R$ is the radius parameter and $a$ the skin diffuseness, and $\theta$ is the polar angle. A random rotation is applied after the sampling process. The spherical harmonic functions $Y_l^m(\theta)$ and the parameters $\beta_i$ account for the possible deformation from a spherical shape. The default Woods-Saxon parameters for uranium are $\beta_2=0.28$, $\beta_3=0$, $\beta_4=0.093$, $a=0.55$ fm, and $R=6.81$ fm \cite{Filip:2009zz, Masui:2009qk,Hirano:2012kj,Shen:2014vra,Schenke:2014tga,Schenke:2020mbo}.   Following~\cite{Moreland:2014oya,Schenke:2020mbo}, we further impose a minimal distance of $d_{\rm min}=0.9\,{\rm fm}$ between nucleons when sampling in three dimensions.\footnote{When a nucleon is added and violates the minimum distance criterion with one or more already sampled nucleons, we resample its azimuthal angle $\phi$   to keep the distributions of radial distances and polar angles unchanged \cite{Moreland:2014oya}.} 

We also study smaller nuclei below. For the case of the nucleon density distribution of ${{\rm ^{20}Ne}}$, we use results from the {\it{ab initio}} Projected Generator Coordinate Method (PGCM) \cite{Frosini:2021fjf,Frosini:2021sxj,Frosini:2021ddm,Bally:2023d}. 
We also compare to the case of a spherical ${{\rm ^{20}Ne}}$ nucleus described by a Woods-Saxon distribution with parameters obtained in low energy electron-nucleus scattering \cite{DeVries:1987atn}. In this case the parameters are the radius $R = 2.8$ fm, and skin depth $a = 0.57$ fm.
For ${{\rm ^{16}O}}$ we employ the nucleon density distribution used in \cite{Loizides:2014vua}, which is obtained from a variational Monte-Carlo method (VMC) using the Argonne v18 (AV18) two-nucleon potential+UIX interactions \cite{Carlson:1997qn}.

\begin{figure}
      \includegraphics[width=0.9\columnwidth]{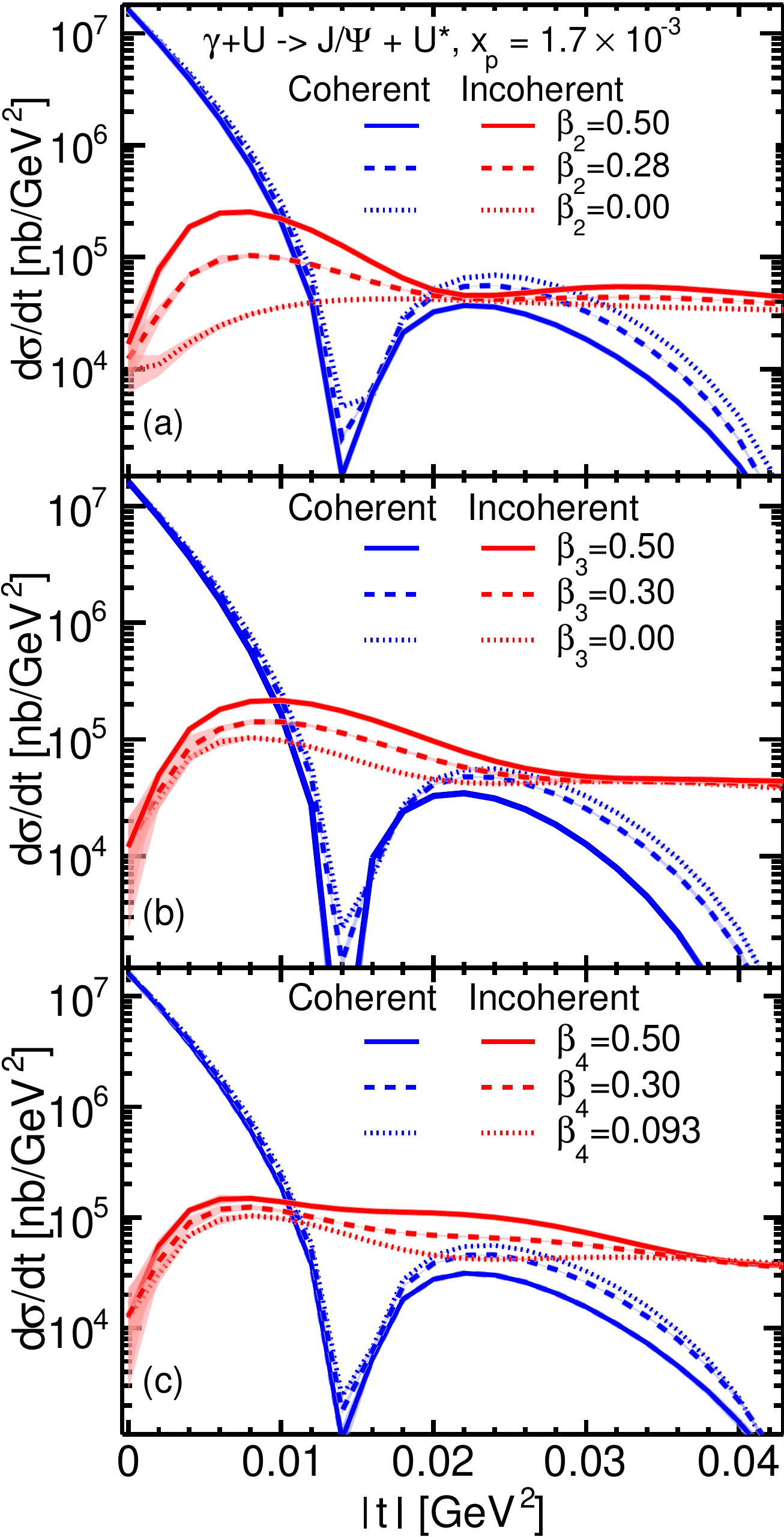}
    \caption{Coherent and incoherent \jpsi photoproduction cross sections at $x_p=1.7\times 10^{-3}$ in e+U collisions for different $\beta_2$ (a), $\beta_3$ (b) and $\beta_4$ (c) values. The bands show the statistical uncertainty of the calculation. } 
    \label{fig:eU_coh_incoh_beta234}
\end{figure}
\begin{figure}
    \centering
    \includegraphics[width=\columnwidth]{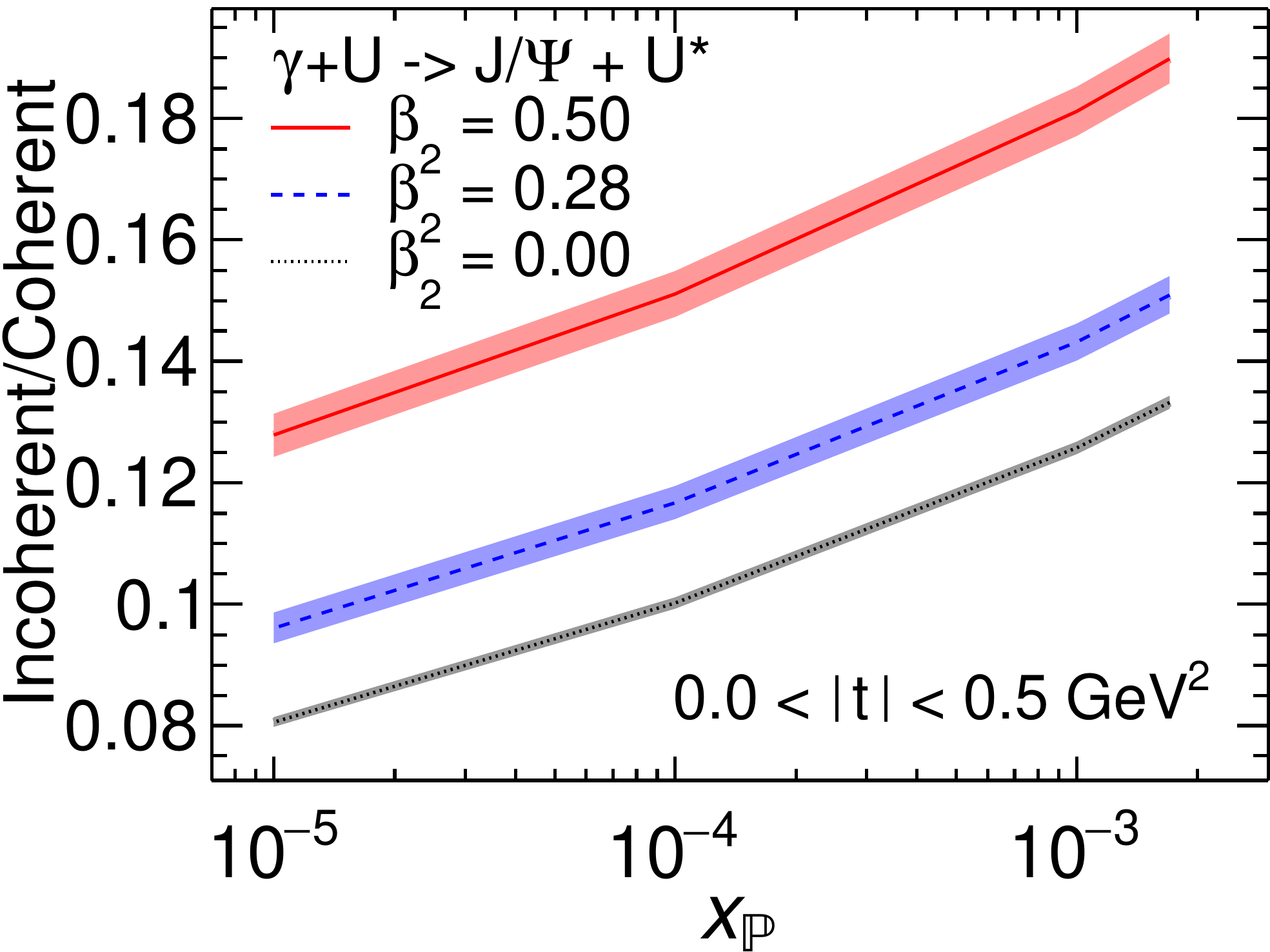}
    \caption{Incoherent-to-coherent \jpsi photoproduction cross section ratio as a function of $\xpom$ in e+U collisions at different initial $\beta_2$ values.  }
    \label{fig:eU_JIMWLK_ratio}
\end{figure}

\begin{figure}
    \centering
    \includegraphics[width=\columnwidth]{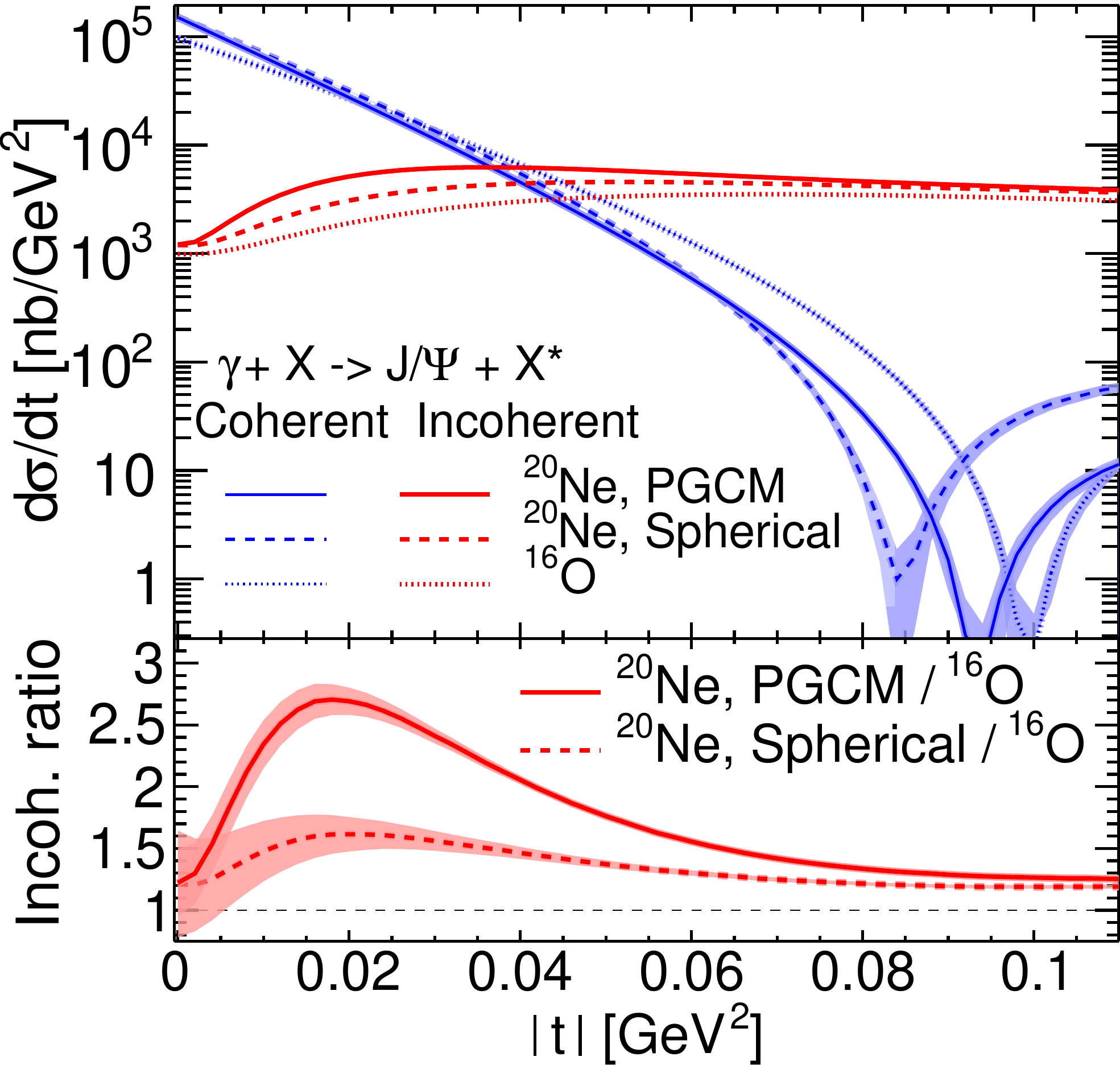}
    \caption{Coherent and incoherent \jpsi photoproduction cross sections in e+$^{20}{\rm Ne}$ and e+ $^{16}{\rm O}$ collisions (top panel), and the incoherent cross section ratios between $^{20}{\rm Ne}$ and $^{16}{\rm O}$ (bottom panel).}
    \label{fig:Ne_O}
\end{figure}

\bigskip
\noindent {\it 3. Sensitivity of exclusive scattering to nuclear deformations.}
Eq.~\eqref{eq:coherent} shows that the coherent cross section is sensitive to the average scattering amplitude 
and as such probes the average structure of the target. The incoherent cross section, Eq.~\eqref{eq:incoherent}, measures the scattering amplitude fluctuations between the different possible color charge configurations. Measuring the total momentum transfer $\Deltat$ allows to constrain the geometry fluctuations in the target at  length scales $\sim 1/\sqrt{-t}$. To determine how sensitive the future EIC measurements are to the nuclear deformations in the currently unexplored small-$x$ region, we vary one of the Woods-Saxon deformation parameters $\beta_2$, $\beta_3$ or $\beta_4$ of uranium, while keeping the others set to their default values.

Figure \ref{fig:eU_coh_incoh_beta234} shows the effect of different levels of quadrupole ($\beta_2$), octupole ($\beta_3$) and hexadecapole ($\beta_4$) deformation on the coherent and incoherent $\gamma^* + \mathrm{U} \to \jpsim + \mathrm{U}^{(*)}$ cross sections for a photon virtuality of $Q^2=0\,\mathrm{GeV}^2$. Increasing the degree of deformation leads to increasing incoherent cross sections.
For example, with a realistic $\beta_2 = 0.28$ the incoherent cross section at $\vert t\vert\sim 0.01~{\rm GeV^2}$ is increased by a factor $\sim 5$ compared to the case with an assumed spherical uranium target with $\beta_2=0.0$. 
This large enhancement is a result of the random orientations of the deformed nucleus in the laboratory frame leading to larger density fluctuations of the  configurations projected onto the transverse plane, which causes larger event-by-event fluctuations in the scattering amplitude. 

Importantly, the different types of deformations manifest in different $\vert t\vert$ regions and can as such be probed separately at the EIC.
The $\beta_2$ modifies the incoherent cross section in the smallest $\vert t\vert \lesssim 0.015~\mathrm{GeV}^2$ region. On the other hand, $\beta_3$ is important in the range $0.005~\mathrm{GeV}^2 \lesssim |t| \lesssim 0.025~\mathrm{GeV}^2$, and $\beta_4$ for $0.015~\mathrm{GeV}^2\lesssim \vert t\vert \lesssim 0.035~\mathrm{GeV}^2$. These effects take place in different $|t|$ ranges because, for example, the quadrupole deformation $\beta_2$ controls the geometric deformation of the target at the longest length scale among these three deformation parameters, which translates into the smallest $\vert t\vert$ region in momentum space.

The  slope of the coherent cross section at low $|t|$ and the position of the first diffractive minimum are not significantly modified by the deformations, which means that the average size of the uranium nucleus is not modified. However, the deformations do affect the coherent spectrum, especially at higher $|t|$. This is due to the fact that (the projection of) the average transverse density profile is different from that of the spherical nucleus when we have non-zero deformation parameters $\beta_i$. Consequently, in addition to the incoherent cross section, the coherent cross section in the relatively large $|t|$ region (after the first diffractive minimum) can be used to access deformations at the EIC.

In order to justify the use of deformation parameters extracted in low energy experiments to the situation of high energy scattering, and in particular to see whether the nuclear deformations are washed out at small-$x$, we apply the perturbative JIMWLK evolution equation~\cite{Mueller:2001uk} as in Refs.~\cite{Mantysaari:2018zdd,Mantysaari:2022sux} to describe the Bjorken-$x$ dependence of the uranium  structure.  
The uranium configurations at the initial $\xpom=1.7\cdot 10^{-3}$ are generated using the same three different $\beta_2$ values as above, and default values are used for $\beta_3$ and $\beta_4$.
Figure~\ref{fig:eU_JIMWLK_ratio} shows the incoherent-to-coherent cross section ratio as a function of $\xpom$, with  the total (in)coherent cross sections integrated within $0.0<\vert t \vert < 0.5~{\rm {GeV^2}}$. This ratio effectively suppresses the uncertainties originating from the modeling of the \jpsi wave function~\cite{Mantysaari:2017dwh}.
The decreasing ratio towards small $\xpom$ implies that the incoherent cross section grows more slowly than the coherent one with increasing energy, because the event-by-event fluctuations are reduced by the evolution~\cite{Mantysaari:2018zdd} (see also~\cite{Cepila:2017nef,Mantysaari:2022sux}).
The difference in the cross section ratio between the different initial quadrupole deformations remains similar throughout the evolution.  We conclude that the fluctuations in the nuclear geometry originating from the deformed structure are not washed out by the JIMWLK evolution, and as such we expect the deformations previously inferred from low-energy experiments to also be visible in high-energy electron-ion collisions at the EIC.

Let us next demonstrate the possibility to probe deformations in the high-energy structure of light nuclei at the EIC, focusing on ${\rm {^{20}Ne}}$ and  ${\rm {^{16}O}}$ (see also~\cite{Mantysaari:2019jhh} where deuteron and helium were considered). 
Figure~\ref{fig:Ne_O} shows both the coherent and incoherent \jpsi photoproduction cross sections at $\xpom=1.7\cdot 10^{-3}$ off a ${\rm {^{20}Ne}}$ nucleus computed from the ab initio PGCM method \cite{Frosini:2021fjf,Frosini:2021sxj,Frosini:2021ddm}, which resembles the shape of a bowling pin. They are compared to the case where we neglect all deformations and use a Woods-Saxon distribution as well as to the case of a ${\rm {^{16}O}}$ target, described as discussed above.

Consistent with the uranium case shown in Fig.~\ref{fig:eU_coh_incoh_beta234}, the bowling pin like shape deformations enhance the incoherent cross section when the PGCM ${\rm {^{20}Ne}}$ is compared to the spherical one at small $\vert t \vert \lesssim 0.05~\mathrm{GeV}^2$. To make the enhancement more visible, the bottom panel of Fig.\,\ref{fig:Ne_O} shows the ratio of the incoherent cross sections for both ${\rm {^{20}Ne}}$ cases to the ${\rm {^{16}O}}$ case. We find that in case of the PGCM ${\rm {^{20}Ne}}$ target, there is an enhancement of up to a factor 2.7, compared to 1.5 when comparing spherical neon to oxygen. The reason for the enhancement of spherical neon compared to oxygen may simply be the larger size of ${\rm {^{20}Ne}}$, which is also visible in the locations of the first diffractive minima in the coherent spectra. The additional enhancement for the case of the PGCM ${\rm {^{20}Ne}}$ nucleus is a result of its bowling pin shape, which essentially results from adding a fifth alpha cluster to the four clusters in ${\rm {^{16}O}}$. 

The largest enhancement is in the regime where the coherent cross section dominates, however, even for $|t|\approx 0.05$ GeV$^2$ where the incoherent cross section begins to dominate the incoherent cross section of ${\rm {^{20}Ne}}$ is almost a factor of 2 larger than that of ${\rm {^{16}O}}$.
At high $|t|$, the incoherent cross sections only differ by a factor $20/16$ originating from the different nuclear mass numbers that affect the overall normalization. This means that the short distance scale fluctuations are identical in oxygen and neon, as it was assumed when constructing the model.

If such measurements can be performed with enough precision at the EIC, they will provide the most direct access to the structure of nuclei over all relevant length scales, from nucleus to subnucleon size scales. 

\bigskip
\noindent {\it 4. Summary. }
We have demonstrated that the coherent and incoherent exclusive vector meson production measurements in $e+A$ collisions are affected by the deformed structure of light and heavy ions. In particular, we have shown that quadrupole, octupole, and hexadecapole deformations can significantly modify the incoherent \jpsi production cross section. Different deformations, on different length scales, affect different regions of momentum transfer, and as such can be explored separately at the EIC. 
We also used numerical solutions to the JIMWLK equation to describe the evolution of the uranium fluctuating structure with decreasing $\xpom$, and found that the evolution suppresses the incoherent cross section at high energies, but does not significantly reduce the effects of the initial deformations.

The comparison between the ${\rm {^{20}Ne}}$ nucleus obtained in modern PGCM calculations and the spherical ${\rm {^{20}Ne}}$ or ${\rm {^{16}O}}$ shows that for light ions a non-trivial shape also results in enhanced incoherent cross sections at low $|t|$. The strongest effect was observed in a $|t|$ region where the coherent cross section dominates, but even in the range where the incoherent cross section is dominant, a factor of 2 enhancement was observed between PGCM ${\rm {^{20}Ne}}$ and ${\rm {^{16}O}}$.
We conclude that correctly predicting incoherent cross sections for the EIC, even within factors of 3 or more,  requires the consideration and possibly even precision calculation of the nuclear deformation of the studied target.
Even more importantly, the predicted large sensitivity of the incoherent cross section to deformations at low $|t|$, along with the previously observed sensitivity to the structure at nucleon and subnucleon size scales at higher $|t|$, implies that this observable carries an unmatched amount of information on nuclear structure over the entire range of relevant size scales. 

This information will be complementary to that obtained from low energy nuclear structure experiments. It will further have direct applications in heavy ion collisions at RHIC, LHC, and other future facilities.

\bigskip
\noindent {\it{Acknowledgments.}}
We acknowledge Benjamin Bally, Thomas Duguet, Jean-Paul Ebran, Mikael Frosini, Giuliano Giacalone, Govert Nijs, Tomás Rodriguez, Vittorio Somà, Wilke van der Schee for kindly sharing unpublished results on the ground-state density of ${\rm{^{20}Ne}}$. We also thank Giuliano Giacalone for useful discussions.
B.P.S. and C.S. are supported by the U.S. Department of Energy, Office of Science, Office of Nuclear Physics, under DOE Contract No.~DE-SC0012704 and Award No.~DE-SC0021969, respectively.  C.S. acknowledges a DOE Office of Science Early Career Award. This material is based upon work supported by the U.S. Department of Energy, Office of Science, Office of Nuclear Physics, within the framework of the Saturated Glue (SURGE) Topical Theory Collaboration.
H.M. is supported by the Academy of Finland, the Centre of Excellence in Quark Matter, and projects 338263 and 346567, and under the European Union’s Horizon 2020 research and innovation programme by the European Research Council (ERC, grant agreement No. ERC-2018-ADG-835105 YoctoLHC) and by the STRONG-2020 project (grant agreement No 824093).
W.B.Z. is supported by the National Science Foundation (NSF) under grant number ACI-2004571 within the framework of the XSCAPE project of the JETSCAPE collaboration.
The content of this article does not reflect the official opinion of the European Union and responsibility for the information and views expressed therein lies entirely with the authors.
This research was done using resources provided by the Open Science Grid (OSG)~\cite{Pordes:2007zzb, Sfiligoi:2009cct}, which is supported by the National Science Foundation award \#2030508.

\bibliography{refs}

\end{document}